\documentstyle[epsfig,longtable]{aipproc}

\begin{document}
\underline{Proceedings: After the Dark Ages -9th October Conference,
College Park, 1998}

\title{Do Small-Scale Dark Matter Fluctuations Govern the
Fragmentation of Primordial Gas?}
 
\author{Volker Bromm, Paolo Coppi, and Richard Larson} {
\address{Department of Astronomy, Yale University, New Haven, CT 06520-8101}

\maketitle

\begin{abstract}
In order to constrain the initial mass function of
the first generation of stars (Population~III), one
has to study the fragmentation properties of primordial gas.
We present results from 3D simulations, based on Smoothed
Particle Hydrodynamics, which explore the idea that small-scale
fluctuations in the (cold) dark matter recreate a filamentary and clumpy
structure in the gas component on scales smaller than the initial Jeans 
mass, where all primordial fluctuations would have been wiped out.
\end{abstract}

\section*{Introduction}

In order to ascertain the influence of the very first generation of
stars (Population~III) on cosmology, one has to address the problem
of how primordial (pure H/He) gas collapses and fragments. Whether the
outcome of this process is a supermassive object or a cluster of
lower-mass stars, is determined by the Population~III initial mass
function (IMF), which might differ substantially from the present-day
one. The fragmentation properties depend crucially on the adopted
initial conditions for the cloud collapse. In principle, these are given
by the underlying model of cosmic structure formation. It is straightforward
to specify the global properties of the gas cloud in terms of its average
temperature, density, and chemical abundances \cite{teg}.

Much less well-defined are the density and velocity fields inside the
cloud. The nature of these initial perturbations could govern the ensuing
fragmentation of the gas. There is no direct connection, however, between them
and the fundamental physics which determines the primordial power spectrum
of density perturbations, since primordial fluctuations in the baryon component
below the Jeans length have been wiped out by pressure forces, as described
by the small-scale cutoff in the baryon power spectrum:
\begin{equation} \label{power-spectrum}
|\delta_{\tt{B}}(k)|^{2}=\frac{|\delta_{\tt{DM}}(k)|^{2}}{
(1+\gamma k^{2})^{2}}{\mbox{\ \ \ \ ,}}
\end{equation}
where $|\delta_{\tt{B,DM}}(k)|^{2}$ are the power spectra in the baryon
and dark matter (DM) components, respectively, $k$ is the comoving wavenumber,
and $\gamma^{\frac{1}{2}}\simeq 1 h^{-1} {\tt{kpc}}$ is the comoving 
Jeans length (cf.\cite{peeb}).

In this paper, we investigate whether the small-scale fluctuations
in the (cold) dark matter component, which have not been erased, do govern
the fragmentation of the baryons, which might start with a completely
smooth distribution. At first, the baryons would not be able to follow
the growing DM clumping, due to the opposing effect of pressure forces.
In the course of a nearly isothermal collapse, however, the Jeans mass
decreases as $M_{{\tt{J}}}\propto \rho^{-\frac{1}{2}}$, allowing the
baryons to fall into the resulting DM condensations.

In the following, we present first results on how the baryons
evolve in this scenario.

\section*{SIMULATIONS}

For our 3D hydrodynamics/dark matter simulations, we
use a variant of TREESPH \cite{hk}, into
which we have incorporated the relevant chemistry of
the formation and
destruction of molecular hydrogen, which is the main coolant below $10^{4}$ K
in the absence of metals. We use an improved version of the H$_{2}$ cooling
function which takes into account quantum effects at low temperatures
($T<600$ K)\cite{gp}.
In order to follow the evolution well into the regime of highly developed 
clumping, we have devised an algorithm to merge high-density particles
(corresponding to excessively small timesteps) into more massive ones,
which enables us to follow the evolution
beyond the point where otherwise the Courant-condition would force
the calculation to a halt. The merging mechanism
allows the high-density particles to
continually accrete nearby particles, thereby modelling the physics of
accretion and merging in an approximate way. 

Our starting model consists of a spherical configuration with a total
mass of $M=10^{6}$ M$_{\odot}$, a radius of $R=80$ {\tt{pc}}, and a
spin parameter of $\lambda=0.05$. In CDM-like scenarios, these are 
typical values for a $3\sigma$ peak, virializing at $z\simeq 30$ \cite{teg}.
On top of the originally homogeneous DM mass distribution, we imprint
fluctuations by assigning initial velocities to the DM particles,
as prescribed by the Zel'dovich approximation \cite{zel}.
The Zel'dovich velocity field is calculated with a power spectrum
$|\delta_{\tt{DM}}(k)|^{2}=A k^{-2.9}$, corresponding to the small-scale end
of the standard CDM spectrum. The amplitude $A$ is chosen to match
$\sigma_{M=10^{6}M_{\odot}}=0.5$ at $z\simeq 30$, appropriate for a collapsing
$3\sigma$ peak.
Embedded in the DM halo is a homogeneous gas cloud with $\Omega_{{\tt{B}}}=
0.05$, an initially isothermal $T_{i}=1000$~K, an H$_{2}$ fraction
of $10^{-3}$, and a free electron abundance of $10^{-4}$ (cf.\cite{teg}).
With the exception of the always present SPH shot-noise ($\propto 
N^{-\frac{1}{2}}$), there are no density fluctuations in the baryonic
component.

We have performed the simulations with $N=65536$ particles in each component,
as well as a comparison calculation at low resolution ($N=8192$).
Fig.~\ref{myfirstfigure} shows the DM and baryon distribution after
one free-fall time. The collapsing dark matter has been organized into
a pronounced structure of filaments and clumps, in response
to the initial Zel'dovich
velocity field. The cooling baryons have begun to condense into the
DM troughs, closely following the DM morphology. Consequently, gas fragmentation
is induced before the DM fluctuations are washed out by the process of
violent relaxation, which will eventually lead to a smooth (roughly
isothermal) mass distribution.

\begin{figure} 
\centerline{\epsfig{file=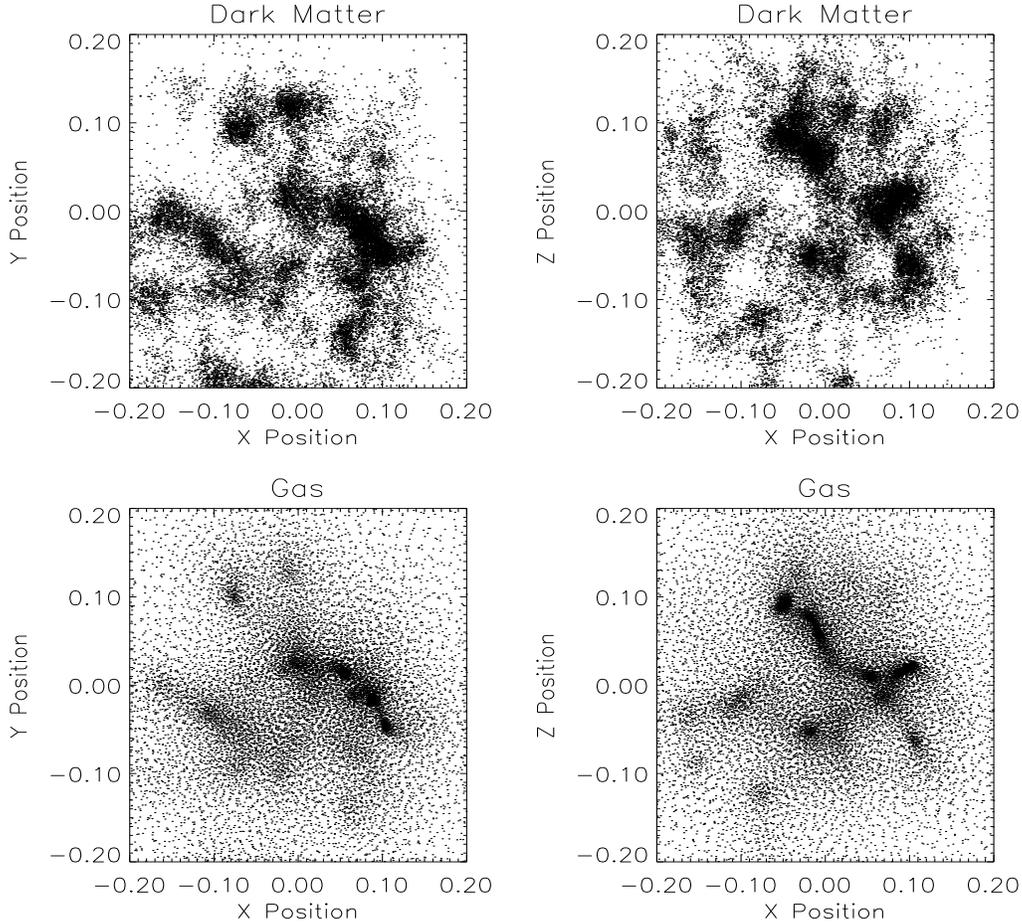,width=14.0cm,height=12.6cm}}
\vspace{10pt}
\caption{
{\it Left column}: Face-on view. {\it Right column}: Edge-on view.
Dark matter and baryon distribution after one free-fall time.
$N_{\tt{SPH}}=N_{\tt{DM}}=65536$. 
Shown is a blow-up of the central region, where
a dimensionless length of 0.1 corresponds to $8 {\tt{pc}}$. The baryons
start to fall into the DM troughs.
}\label{myfirstfigure}
\end{figure}
In Fig.~\ref{mysecondfigure}, we present the distribution of gas at a 
slightly later time, $t=1.2 t_{\tt{ff}}$, for the low-resolution run. Here, most
of the gas ($\sim 60\%$) resides in high-density clumps, which are a result
of the merging procedure. 
The resulting mass spectrum
of these merged clumps follows a power-law with roughly the Salpeter slope.
This result should be taken {\it cum grano salis}, and its robustness has
to be tested in runs with higher resolution and with different initial
conditions.

\begin{figure} 
\centerline{\epsfig{file=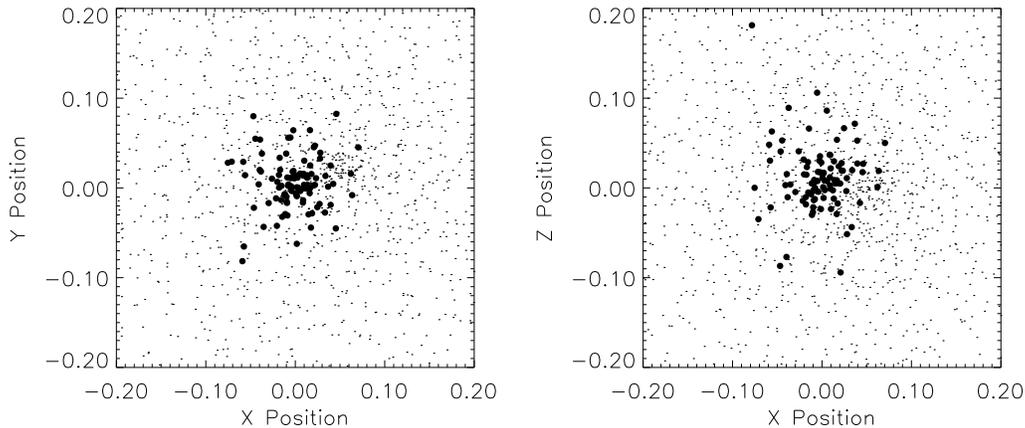,width=14.0cm,height=6.4cm}}
\vspace{10pt}
\caption{
Gas distribution after $t=1.2t_{{\tt{ff}}}$.$N_{{\tt{SPH}}}=N_{\tt{DM}}=8192$. 
Same units as in Fig.~\ref{myfirstfigure}. Most of the gas has been 
incorporated into high-density clumps 
({\it heavy dots}).
}\label{mysecondfigure}
\end{figure}

\section*{OUTLOOK}

Although idealized, the appeal of the above scenario lies in the promise of
being able to specify the initial conditions of the primordial star
formation problem in a non-arbitrary way, directly connected to the
underlying model of cosmic structure formation. But there remain many
caveats which have to be explored. Does it make sense to treat a
high-$\sigma$ peak in isolation? Is the initial baryon distribution
really smooth, or instead structured through complex processes which
are difficult to specify without simulating a much larger region of the
Universe?

We hope to gain a better understanding of these issues by comparing
our results with those from large-scale cosmological simulations. 
Independent of the
particular model proposed above, it
makes sense to study the fragmentation of primordial gas in a broad range
of circumstances (e.g., using various spectral indices and amplitudes for both
the baryon and DM components). Currently, we are undertaking such a
comprehensive study, on which we will report elsewhere.

Support from the NASA ATP grant
NAG5-7074 is gratefully acknowledged.

\end{document}